\documentclass[aps,amsfonts,prl,twocolumn,showpacs, superscriptaddress]{revtex4-1}

\usepackage{subfigure}
\usepackage{textcomp}
\usepackage{graphicx}
\usepackage{amssymb}
\usepackage{amsmath}
\usepackage{bm}
\usepackage{amsmath, amsthm, amssymb}
\usepackage{dsfont}
\usepackage[colorlinks=true,linkcolor=blue,urlcolor=blue,citecolor=blue]{hyperref}
\usepackage{multirow}
\usepackage{url}

\usepackage{pdfpages} % include appendix
\makeatletter
\AtBeginDocument{\let\LS@rot\@undefined}
\makeatother

\def\bra#1{\mathinner{\langle{#1}|}}
\def\ket#1{\mathinner{|{#1}\rangle}}

\def\beq{\begin{equation}}
\def\eeq{\end{equation}}
\def\bea{\begin{eqnarray}}
\def\eea{\end{eqnarray}}

\begin{document}

\title{Asymptotically exact theory for nonlinear spectroscopy of random quantum magnets}

\author{S. A. Parameswaran}
\affiliation{Rudolf Peierls Center for Theoretical Physics, Clarendon Laboratory, University of Oxford, Oxford OX1 3PU, UK}
\author{S. Gopalakrishnan}
\affiliation{Department of Physics and Astronomy, CUNY College of Staten Island, Staten Island, NY 10314}
\affiliation{Physics Program and Initiative for the Theoretical Sciences, The Graduate Center, CUNY, New York, NY 10016, USA}

\begin{abstract}
We study nonlinear response in quantum spin systems {near infinite-randomness critical points}. Nonlinear dynamical probes, such as two-dimensional (2D) coherent spectroscopy, can diagnose the nearly localized character of excitations in such systems. {We present exact results for nonlinear response in the 1D random transverse-field Ising model, from which we extract information about critical behavior that is absent in linear response. Our analysis yields exact scaling forms for the distribution functions of relaxation times that result from realistic channels for dissipation in random magnets}. 
We argue that our results capture the scaling of relaxation times and nonlinear response in generic random quantum magnets in any spatial dimension.
\end{abstract}

\maketitle
The use of strong electromagnetic fields to probe solid-state systems and pump them into exotic states has been a fruitful research direction. One paradigm for these experiments has been ``pump-probe spectroscopy,'' where a system is pumped with an intense field, creating a far-from equilibrium state, whose response to a weaker ``probe'' field is subsequently measured~\cite{PhysRevLett.87.237401, PhysRevLett.85.2204, PhysRevLett.97.067402, PhysRevB.78.245113, PhysRevB.78.205119, thorsmolle2010ultrafast, fausti2011light, wang2013observation, mitrano2016possible, fischer2016invited, babadi1}. 
Two-dimensional coherent spectroscopy (2DCS)~\cite{Mukamel1995, hamm2005principles, axt2004femtosecond, cundiff2013optical, woerner2013ultrafast} is a conceptually similar multi-pulse technique, but operates in a regime where the pump changes the state of the system only weakly. Instead of creating and characterizing far-from-equilibrium states, 2DCS probes multitime correlation functions of a given equilibrium state, which contain qualitative information not captured by linear response: e.g., they can distinguish between ``inhomogeneous'' broadening (i.e., a continuum of response due to many separate sharp modes with wide frequency spread) and ``homogeneous'' broadening (i.e., broadening due to finite excitation lifetimes). Consequently 2DCS can isolate interaction effects in settings where linear response does not diagnose the central phenomena of interest~\cite{PhysRevLett.109.157005, rosenow, sg_heating, kozarzewski, rehn2016, lcks,PhysRevLett.122.220601,PhysRevB.97.214208}, such as in systems exhibiting fractionalization~\cite{wan2019resolving, choi2020theory} or localization~\cite{mahmood2020observation, thorsmolle2010ultrafast, muller2019towards}. 

Here, we construct an asymptotically exact theory for the response of random quantum systems probed using 2DCS (or, more generally, pump-probe spectroscopy).  We obtain exact results near the quantum critical point (QCP) of the random transverse-field Ising model (RTFIM). This is a paradigmatic example of a system controlled by an infinite-randomness fixed point (IRFP)~\cite{ma1979random, fisher1992random, fisher1995critical, motrunich2000infinite}, %
{and hence can be modeled}
as an ensemble of weakly interacting two-level systems (TLSs). The properties of this ensemble are determined by scaling exponents associated with the {IRFP}. Anomalous exponents persist away from the  {QCP} due to Griffiths effects~\cite{fisher1992random}. While some information about the TLS distributions can be extracted from linear response~\cite{mdh, PhysRevLett.106.137202, PhysRevLett.111.147203}, we  
{show} that nonlinear response reconstructs the full TLS distribution function as well as the residual interactions among TLSs.

Our central results concern the lifetimes of TLSs at %
{IRFPs} and in quantum Griffiths phases. %
Although the coupling of the system to its environment is irrelevant in the renormalization group sense, it is dangerously irrelevant, and causes the TLSs to have finite lifetimes. %
For random quantum magnets coupled to phonons (or other non-magnetic baths), we find that the TLS relaxation times are power-law distributed both at criticality and into the Griffiths phase. \emph{Local} probes measure the relaxation of a typical TLS, which is exponential with a rate we compute. However, the \emph{spatially averaged} response probed by most optical experiments picks up the entire broad spectrum of relaxation times. We show that 2DCS response in the frequency-time plane extracts exponents characterizing both the relaxation-time and resonance-frequency distributions. {Although exact analytical results are only possible in 1D, we}
 argue that the phenomenology of the response is generic to a large class of random quantum magnets in any spatial dimension. %, 

\emph{Random TFIM}.-- 
{We begin with the 1D  RTFIM,} 
\beq\label{eq:RTFIMham}
H_{\mathrm{RTFIM}} = - \sum_i (h_i \sigma^z_i + J_i \sigma^x_i \sigma^x_{i+1}),
\eeq
where the $h_i$, $J_i$ are  positive i.i.d. random variables.  
(Swapping $\sigma^x$ and $\sigma^z$  relative to convention yields a more natural TLS basis later.) 
The %IRQCP
{QCP} in this model occurs when $2\delta \equiv (\overline{\ln h_i}  - \overline{\ln J_i}) = 0$, where $\overline{(\cdots)}$ denotes a disorder average.
The RTFIM can be iteratively diagonalized by an asymptotically exact real-space renormalization-group (RSRG) method~\cite{fisher1992random, fisher1995critical, fisher1999phase}. For the ground state, the RSRG rules are as follows. One picks the strongest coupling. If it is a bond $J_i$, one fuses the spins it connects into a superspin, which experiences an effective transverse field $h_{i-1} h_{i}/J_i$. If it is a transverse field $h_i$, one eliminates site $i$ by placing it in its $\ket{+\hat{z}}$ state, creating an effective perturbative coupling $J_i J_{i+1} / h_i$. These steps are iterated until all spins have been decimated.  
{The RSRG yields paramagnetic (PM) and ferromagnetic (FM) phases separated by a QCP, {which is an IRFP} with extremely broad power-law distributions of renormalized couplings {(hence the term ``infinite randomness'').} {This leads to the following properties}: 
\textbf{(1)}~{Critical} spatial and temporal fluctuations are infinitely anisotropic, and scale via the relation $\ln t \sim \sqrt{x(t)}$. 
\textbf{(2)}~At criticality, the typical magnetic moment of a superspin at scale $x$ is $\mu\sim x^{\phi/2}$, where $\phi$ is the Golden mean.
\textbf{(3)}~The {QCP} is flanked by Griffiths phases on both sides. In the PM Griffiths phase, for example, the system as a whole is not magnetically ordered, but has rare FM regions which locally appear to be on the ``wrong side'' of the transition and dominate low-frequency response, as we discuss next. 

\emph{Rare TLSs in the Griffiths phase}.--- %
A field decimation effectively decouples a superspin (FM cluster) from the rest of the system, whereas a bond decimation grows a FM cluster. A cluster that decouples at energy scale $\varepsilon$ contributes to dynamics at $\omega=\varepsilon$, but freezes out at lower frequencies. Clusters that are slow compared to $\omega$ are also unimportant to \emph{dissipative} response%. Thus,
: to understand 
response at $\omega$, we must characterize %
{FM clusters} generated by field decimations occurring at scale $\varepsilon=\omega$.  

In the PM, the system initially looks critical on short distances, but eventually on coarse-graining out to the correlation length $\xi \sim (\ln \varepsilon^*)^2 \sim \delta^{-2}$, the energy scale reduces below $\varepsilon^* \sim e^{-1/\delta}$. The RG then crosses over into the off-critical PM regime where it is overwhelmingly likely to decimate fields rather than bonds. 
For frequencies  $\omega <\varepsilon^*$, the system can therefore be viewed as a set of
{rare FM clusters}  weakly coupled by the residual bond terms. Such clusters 
contribute anomalous power laws to low-frequency response: a locally FM cluster of $l$ sites has an exponentially small probability {(in $l$)} of occurring.
{The two  parity eigenstates $\ket{\pm} = \frac{1}{\sqrt{2}} (\ket{\uparrow}^{\otimes l}\pm  \ket{\downarrow}^{\otimes l})$  of each cluster form a TLS
separated from the other states by an energy gap.} %
 {Tunneling between these occurs at $l^{\text{th}}$ order in perturbation theory, leading to a TLS resonant frequency  that obeys $\log (1/\omega) \propto l$.}
 {Combining this with the exponential-in-$l$ probability, we find that} %
  TLSs    flippable at frequency $\omega$ are of size $\sim |\ln \omega|$, i.e. are only power-law rare in $\omega$, {with spacing $x(t) \sim t^{1/z}$ where $z$ is a continuously varying exponent computed below}.
  {Note that  $x(t)$, the characteristic lengthscale at time $t\sim\omega^{-1}$, is the \emph{spacing} between rare TLSs, and is distinct from the parametrically smaller \emph{size} $l(t)\propto \log t $ of each TLS.} %

{Precise scaling forms result from} running RSRG from microscopic energy scale $\Omega_I$ (set to $1$ throughout) to the
{probe} scale $\Omega$, 
 where the remaining degrees of freedom 
 {are} these rare TLSs. 
 The bond and field distributions are 
 \beq\label{eq:distribution}
P_\Omega(J) =\frac{u_\Omega}{\Omega} \left(\frac{J}\Omega\right)^{u_\Omega-1}, \,\,\,\,\,\rho_\Omega(h) =\frac{\tau_\Omega}{\Omega} \left(\frac{h}\Omega\right)^{\tau_\Omega-1}
 \eeq
 with $u_\Omega = \frac{2\delta}{e^{2\delta\Gamma} -1}$ and $\tau_\Omega = \frac{2\delta}{1 -e^{-2\delta\Gamma}}$, where $\Gamma = \ln\Omega_I/\Omega$. As $\delta\to \Omega$, $u_\Omega\sim\tau_\Omega\sim 1/\Gamma$, and both distributions tend to $P(J) = \frac{1}{\Gamma J} \left( \frac\Omega J\right)^{1-1/\Gamma}$ which broadens as the RG flows to $\Gamma\to \infty$, as is characteristic of IRQCPs.  $\rho_\Omega(\varepsilon)$ is the density of TLSs with splitting  $\varepsilon<\Omega$ at scale $\Omega$. The effective size of a rare TLS at energy scale $\varepsilon$ is obtained by running the RG to $\Omega = \varepsilon$ and then using the rare region arguments above, yielding
 \beq \label{eq:lscaling}
l_\varepsilon = \tau_\varepsilon^{-1}|\ln \varepsilon| \sim  |\ln \varepsilon|({1- \varepsilon^{2\delta}})/{2\delta}.
 \eeq
{The full expression~\eqref{eq:lscaling} captures} both the  $l_\varepsilon \sim |\ln \varepsilon|^2$ scaling at criticality (by taking $\delta\to 0$ before $\varepsilon\to0$) as well as finite-$\delta$ {PM Griffiths} behavior (the opposite order of limits). We may estimate the magnetic moment of a TLS  by viewing it  as composed of $n\sim l/\xi$ critical clusters of size $\xi$ and moment $\mu_\xi\sim \xi^{\phi/2}$, using $\xi\sim \delta^{-2}$, and replacing $\delta$ by its renormalized value: 
  \beq \label{eq:muscaling}
\mu_\varepsilon  \sim |\ln \varepsilon| \left[2\delta/(1 - \varepsilon^{2\delta})\right]^{1-\phi},
 \eeq
which becomes $\mu \sim|\ln \varepsilon|^\phi$ at criticality (as noted above).

\emph{Relaxation processes}.---The RSRG  generates infinitely sharp TLSs, corresponding to strictly localized excitations. In realistic systems, these TLSs eventually relax. Relaxation can occur either because the system is coupled to an extrinsic reservoir, or acts as its own bath~\cite{basko2006metal} because of interactions; we focus on extrinsic baths, but our results should also generalize, with some modifications, to intrinsic baths 
treated self-consistently~\cite{basko2006metal, PhysRevB.89.220201, gn}. Two other key distinctions in terms of relaxation dynamics are: (i)~between magnetic baths that couple directly to the order parameter $\sigma^x$ (and therefore involve magnetic degrees of freedom, e.g., nuclear spins) and non-magnetic baths that do not (e.g., phonons), and (ii)~between baths 
with rapidly vanishing low-energy spectral density $\mathcal{J}(\omega) \sim \omega^s$, $s > 1$ (superohmic baths, defined more precisely below) and those with $s \leq 1$ (i.e., ohmic or subohmic)~\cite{leggett_review}. Our central new results involve non-magnetic baths; before turning to these, we briefly comment on magnetic ones~\cite{mms1, mms2, vojta2003, schehr2006, hoyos2008, hoyos2012}. A magnetic bath always has a matrix element ($\propto \mu_\varepsilon$) to flip a single TLS, and in the ohmic/subohmic cases 
  couples strongly to  TLSs and destroys the IRQCP~\cite{mms1, mms2, vojta2003, schehr2006, hoyos2008, hoyos2012}. For superohmic baths, however, the IRQCP survives, and one can straightforwardly compute excitation lifetimes using Fermi's Golden Rule, as $\tau^{-1} \sim \mu_\varepsilon^2 \varepsilon^s$. This sets timescales for both energy relaxation ($T_1$) and dephasing ($T_2$).

\begin{figure}[t!]
\includegraphics[width=\columnwidth]{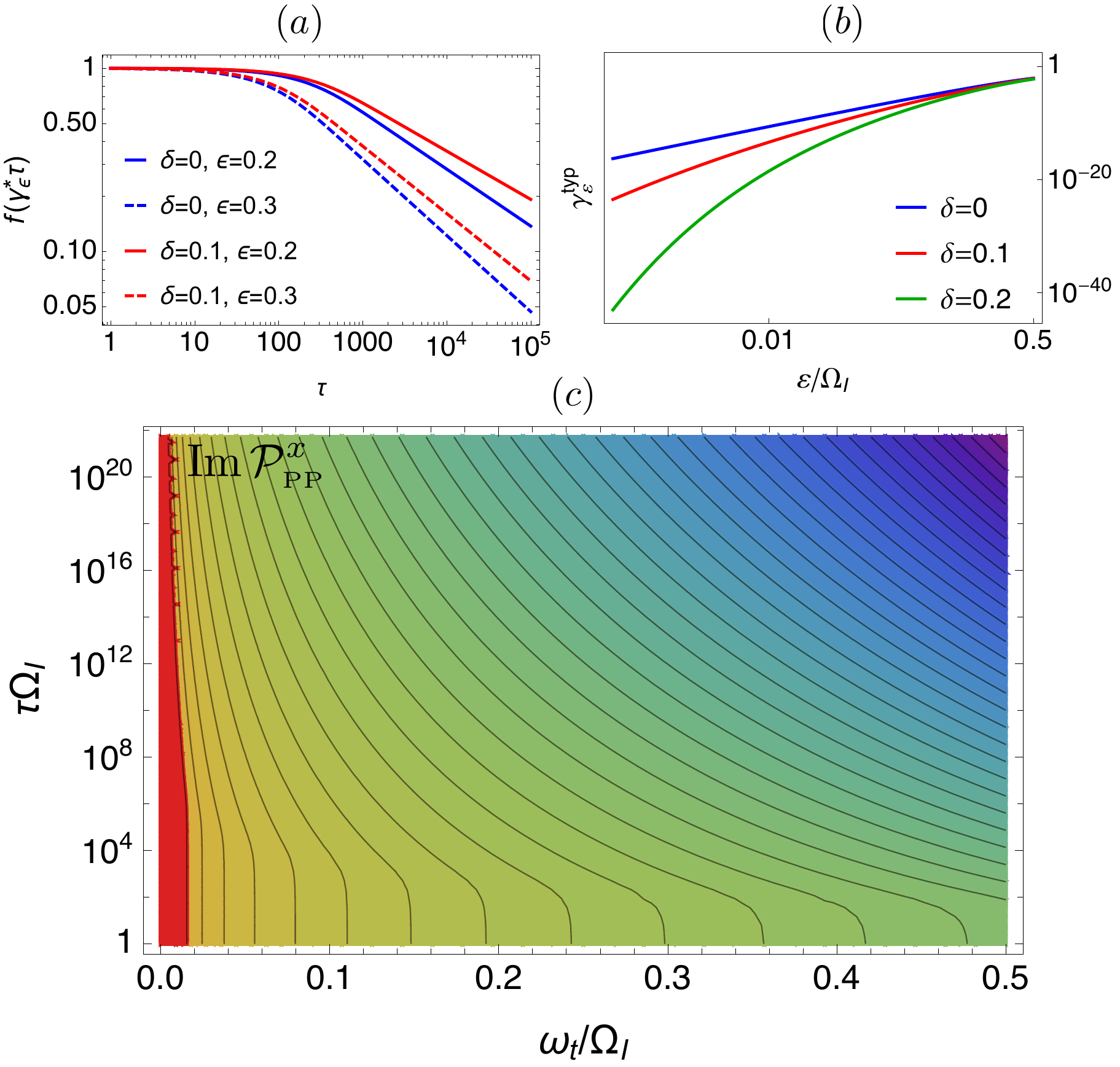}
\caption{\label{fig:combined} Relaxation and 2DCS in the RTFIM. (a) The average decay is power-law, and is faster at the QCP (blue) than in the PM Griffiths phase (red); decay is slower at lower frequencies (solid vs. dashed lines). However, evolution to $\delta=0$ is smooth. (b)  The \textit{typical} decay is exponential, $e^{-\gamma^{\text{typ}}_\varepsilon \tau}$, and the $\varepsilon$-dependence of   $\gamma^{\text{typ}}_\varepsilon$ at QCP is a power-law with logarithmic corrections, sharply distinct from its stretched-exponential behaviour in the Griffiths phase. (c)``Mixed'' 2DCS plot of the average pump-probe  response of the RTFIM at criticality, Eq.~\eqref{eq:2DCSPP}. Contour lines and color scale are logarithmic. The 2DCS response for $\delta>0$ is similar and evolves smoothly out of $\delta=0$. (We used a superohmic  $s=2$ bath.)}
\end{figure}

We now turn to the more delicate relaxation channel due to the coupling of TLSs to phonons, which modulate the distance 
 (hence coupling) between nearby spins. We can incorporate phonons through the change $J_i \mapsto J_i (1 + \hat{X}_i)$ to the bond terms in~\eqref{eq:RTFIMham}, where $\hat{X}_i = \sum_{i} \lambda_{i} (b^\dagger_{i} + b_{i})$ is the coupling to phonon modes, treated as purely harmonic~\cite{leggett_review} with Hamiltonian $H_b = \sum_i \Omega_i b^\dagger_i b_i$, 
 For now we also treat each spin as coupled to its own phonon bath; 
later we comment on {more realistic cases} 
We introduce the spectral density of the bath, $\mathcal{J}_i(\Omega) =\mathcal{J}(\Omega) = \pi \sum_{\alpha} \lambda_{i,\alpha}^2\delta(\Omega -\Omega_{i,\alpha}) \equiv g \Omega_c^{1-s} \Omega^s e^{-\Omega/\Omega_c}$, where $g$ is a dimensionless measure of dissipation and $\Omega_c$ is a high-frequency cutoff for the bath~\cite{leggett_review}. 
 Due to the paucity of  $\Omega\to 0$ bath modes,
  low-frequency TLSs are always weakly coupled to a %the
  superohmic bath, so the RG proceeds as  in the closed system. Since each bond is coupled to its own bath, 
  a TLS at energy $\varepsilon$ is coupled to $O(l_\varepsilon)$ 
different baths, so the effective 
 spectral density is $\sim l_\varepsilon \mathcal{J}$.

Since phonons transform trivially under the Ising symmetry, the bath cannot couple to the order parameter; instead, it can only couple \emph{diagonally} to the TLS, i.e., it can modulate its transverse field, leading to pure dephasing. A TLS subject to  superohmic pure dephasing  
remains phase-coherent  to 
infinite time~\cite{PhysRevLett.109.233601}. This is because a superohmic bath has little weight at low frequencies, so does not cause long-time drift of the TLS resonant frequency. (For similar reasons, 3D crystals have sharp Bragg peaks despite the presence of phonons.)
To get true broadening of the TLS line in the zero-temperature limit, we therefore need to consider 
longitudinal relaxation processes where a putatively decoupled TLS flips its state via the weak residual bonds that couple it to other, lower-energy TLSs.  The role of the bath is to place this decay channel on-shell.

To compute the relaxation rate, consider resonantly exciting a TLS with splitting $\varepsilon$. Although it has nominally decoupled at that energy scale, it still has residual couplings $J_\varepsilon$ to its (lower-energy) neighbors. It can therefore decay into the bath via a process where it ``flip-flops'' with its nearest neighbors (at energy $\varepsilon' \ll \varepsilon$) while depositing 
energy $\varepsilon-\varepsilon'$ into the bath. Such decay occurring at rate $\gamma$ yields relaxation times $T_2 = 2T_1 = \gamma^{-1}$. The  rate of decay of a TLS of energy $\varepsilon$ via flip-flop processes with another TLS with energy $\varepsilon<\varepsilon'$ can be estimated via Fermi's Golden Rule to be $\gamma_{\varepsilon, \varepsilon'} \sim l_\varepsilon J_\varepsilon^2\mathcal{J}(\varepsilon -\varepsilon') \sim  gJ_\varepsilon^2\frac{|\ln \varepsilon|}{\tau_\varepsilon} (\varepsilon-\varepsilon')^s$, where $l_\varepsilon$ accounts for the enhanced bath spectral density. Since $J$ is broadly distributed, so is {$\gamma$}. . 
We compute the distribution of $\gamma$ in terms of the distributions of the TLSs  
with $\varepsilon'<\varepsilon$ and the residual couplings obtained by running RSRG down to scale $\varepsilon$, using \eqref{eq:distribution}, 
{and neglecting the weak $\varepsilon'$-dependence of $\gamma_{\varepsilon, \varepsilon'}$}:
 \begin{eqnarray}\label{eq:gammadist}
 P_\varepsilon(\gamma) &=& \int_0^\varepsilon d\varepsilon'\, \rho_\varepsilon(\varepsilon') \int_0^\varepsilon dJ P_\varepsilon(J)\delta(\gamma - \gamma_{\varepsilon,\varepsilon'})\nonumber\\
 &=& \frac{u_\varepsilon}{2 \gamma} \left(\frac{\gamma}{\gamma^*_\varepsilon}\right)^{\frac{u_\varepsilon }{2}}\Theta(\gamma^c_\varepsilon -\gamma),
 \end{eqnarray}
where  $\gamma^*_\varepsilon \sim l_\varepsilon \varepsilon^{s+2} \sim \frac{1- \varepsilon^{2\delta}}{2\delta} |\ln \varepsilon| \varepsilon^{2+s}$. The relaxation rates are broadly distributed; thus, the late-time response at scale $\varepsilon$ {averaged} over the random environments of the TLSs at that scale will be dominated by the tail of $P_\varepsilon(\gamma)$.
This corresponds to TLSs with anomalously weak relaxation, and leads to a slow power-law decay of the signal [Fig.~\ref{fig:combined}(a)]. However, the response to {\it local} probes is sensitive to the {\it typical} decay rate, 
\begin{eqnarray}\label{eq:gammatyp}
\gamma^{\text{typ}}_\varepsilon = e^{\overline{\ln \gamma}} = \gamma^*_\varepsilon e^{-\frac{2}{u_\varepsilon}}
&=&  \frac{1- \varepsilon^{2\delta}}{2\delta} |\ln \varepsilon| \varepsilon^{s+2} e^{\frac{1- \varepsilon^{-2\delta}}{\delta}},
\end{eqnarray}
which has a stretched-exponential suppression in the PM Griffiths phase ($\delta>0$) that is absent at criticality [Fig.~\ref{fig:combined}(b)]. Consequently, local response at energy $\varepsilon$ sharpens on moving off-criticality into the PM.

\emph{2DCS Response}.---We now discuss how to probe these lineshapes using nonlinear spectroscopy, focusing on 2DCS~\cite{Mukamel1995, hamm2005principles, muller2019towards}. We consider the following protocol:  initialize the TLS in its ground state; apply two sharp pulses $A$ and $B$, separated by time $\tau$, that couple to the magnetization; wait time $t$ and measure the magnetization. In terms of the Pauli matrices $\sigma^{x,y,z}$, the TLS has Hamiltonian $H_0 = -\frac{\varepsilon}2\sigma^z$,   $\ket{0} = \ket{+\hat{z}}$ and the pulses are $\delta$-function kicks that couple to $\sigma^x$. {As these rotate the TLS by  $\theta_{A,B}$ in the $yz$ plane}, corresponding to action ${R}(\theta) = e^{i\frac{\theta}{2}\sigma^x}$, {absent relaxation} the protocol prepares the state (setting $\hbar=1$)
$\ket{\psi(t;\tau)} = e^{-i H_0 t} R({\theta_A}) e^{-i H_0 \tau} R({\theta_B})\ket{0},$
and we measure ${\mathcal{P}}^x(t, \tau) = \bra{\psi(t; \tau )}  {\sigma}^x \ket{\psi(t;\tau)}$.
In the perturbative limit, $R_\theta \approx 1 + i \theta \sigma^x + \cdots$, and only odd  powers contribute to $\mathcal{P}^x$. 
 The leading 
nonlinear contributions are cubic in $\theta$. There are several distinct cubic channels; a key insight of 2DCS is that these channels oscillate at different frequencies with respect to $t$ and $\tau$~\cite{hamm2005principles}. 
 We focus on the ``pump-probe'' (PP) contribution, in which both  bra and ket are flipped by the initial pulse,  
{${\mathcal{P}}^x_\text{\tiny PP}\! \propto\! \text{Im}\!\bra{0}\sigma^x e^{i H_0 (\tau + t)} \boldsymbol\sigma^x e^{-i H_0 t} {\sigma}^x e^{-i H_0 \tau}\sigma^x \!\ket{0},
$}
which looks like a linear-response correlator measured in the state $\sigma^x\ket{0}$ driven out of equilibrium by the drive. {When relaxation is included,} the PP response is subject to longitudinal relaxation between pulses, and transverse relaxation after the second pulse, yielding
\beq
{\mathcal{P}}^x_\text{\tiny PP} (t,\tau) \sim e^{-\tau/T_1}e^{-t/T_2} \sin \varepsilon t, \label{eq:PPRrelax} 
\eeq
and hence distinguishes homogeneous broadening due to longitudinal relaxation from inhomogeneous broadening: only the former depends on $\tau$. In a 2DCS experiment the PP response is convolved with 
those from other channels; individual channels are isolated by Fourier-transforming the full response with respect to $(t, \tau)$. 

Near an 
{IRFP} this conventional approach is complicated by the broad distribution of low-frequency TLSs and corresponding relaxation times.  The 2DCS response is neither purely reactive nor purely absorptive; {in practice, to extract the absorptive response one often assumes a Lorentzian lineshape, an assumption that fails for IRFPs}.
We propose the following resolution to this difficulty. Instead of Fourier transforming with respect to both $t, \tau$, we consider the \emph{mixed} response in the $(\omega_t, \tau)$ plane. 
The PP channel can be isolated from other channels by time-averaging in
 $\tau$ 
 over a window $\omega_t^{-1}$. 
To compute the full PP response, we  combine the single-TLS response \eqref{eq:PPRrelax}  with the following: (i) the density of TLSs with splitting $\varepsilon$ is $\rho_\Omega(\varepsilon) \sim 1/\varepsilon^{1-2\delta}$; (ii) such TLSs relax with $T_2 = 2T_1 \sim \gamma^{-1}$ distributed according to \eqref{eq:gammadist}; and (iii)~each TLS couples to a probe field via its moment $\mu_\varepsilon$ \eqref{eq:muscaling}, with 
{${\mathcal{P}}^x_\text{\tiny PP} \sim\sim \mu_\varepsilon^4$.} 
 Accordingly, the averaged mixed pump-probe response for $\omega_t>0$ scales as
\beq\label{eq:2DCSPP}
\!\!\text{Im}\,{\mathcal{P}}^x_\text{\tiny PP} (\omega_t,\tau) \sim \rho_\Omega(\omega_t)
\overline{e^{-\gamma \tau/2}}\sim{\omega_t^{2\delta-1}} {\mu_{\omega_t}^4}f_{\omega_t}(\gamma^*_\varepsilon \tau),
\eeq
where $f_\varepsilon(x) = \frac{u_\varepsilon}{2 x^{u_\varepsilon/2}} \int_0^x \xi^{\frac{u_\varepsilon}{2}-1} e^{-\xi} d\xi$ describes the averaged decay profile at frequency $\varepsilon$ [cf. Fig.~\ref{fig:combined}(a)]. (To arrive at Eq.~\eqref{eq:2DCSPP} we used 
$\gamma_\varepsilon \ll \varepsilon$ to separate out the $\omega_t$ and $\tau$ directions.) Fig~\ref{fig:combined}c shows such a mixed-2DCS portrait of the pump-probe response of the RTFIM at criticality; note the rapid sharpening of the response at low frequencies. A fixed $\omega_t$-slice corresponds to $f_\varepsilon$ and hence allows us to extract $\gamma^*_\varepsilon$, whereas the evolution of the peak height at $\tau=0$ allows us to extract information on the energy-dependent distributions and renormalized moments. (Previous work on anomalous power-law relaxation in 2DCS focused on spectral diffusion~\cite{PhysRevLett.98.080603}, which is distinct from the quenched disorder operational here.)

{\textit{Discussion: beyond the 1D RTFIM}.---Above, we explicitly showed that} nonlinear response {of the 1D RTFIM near criticality} accesses and deconvolves critical data that are only found in certain combinations in linear response: the magnetic moment sets the strength of  nonlinearity, the density of TLSs sets the overall spectral intensity, the lineshape probes the distribution of residual couplings, and the typical excitation lifetimes diagnose whether the system is in the critical or the Griffiths regime. In fact, our results apply more broadly. They immediately generalize to other IRFPs and adjacent Griffiths phases, in any dimension, since these are also described by RSRGs that decouple the system into ensembles of localized few-level systems which transform as irreducible representations of some global symmetry $G$ -- e.g., the QCPs and PM Griffiths phases of 2D RTFIMs and  quantum Potts/clock chains, or the {random-singlet phase of  1D Heisenberg antiferromagnets.}~\cite{ma1979random, XXZ_Fisher_RG, PhysRevLett.76.3001, hyman_yang, motrunich2000infinite}. {Provided  {$G$ is not spontaneously broken and} the bath degrees of freedom transform trivially under $G$, excitations of the system must relax via} the broadly-distributed residual couplings.
Indeed, since all couplings approach the distribution $P(x) \sim 1/x$ at IRFPs, the critical 2DCS response is superuniversal: distinctions between IRFPs appear only in prefactors~\cite{*[{A recent discussion of some aspects of superuniversality in 2D is in
}] [{.}] 2020arXiv200809617K}. {Likewise, the lineshapes in quantum Griffiths phases follow from very general considerations involving the counting of rare regions. At general IRFPs a superspin might respond at multiple discrete frequencies, but since these will be similar in magnitude, this complication does not affect our scaling analysis.}

{One of our key results, the distribution of lineshapes}, depends only on the broad distribution of couplings between low-energy degrees of freedom. 
{This feature is generic} to random quantum magnets in arbitrary spatial dimensions (even if they {lack} an IRFP description), as long as two criteria are satisfied: (i) the low-energy degrees of freedom probed by the response are localized, with residual couplings 
{decaying exponentially in} the spacings between them, which are exponentially distributed; and (ii) {the bath preserves an unbroken global symmetry}. The former behaviour is generic~\cite{bhattlee}, while the latter is typical of several realistic baths, including phonons. 
{However, the density of states (which sets the intensity profile along the $\omega_t$ axis in Fig.~\ref{fig:combined}) and the scaling of {\it typical local} lifetimes are likely model-specific.}

{We took each bond to couple to its own phonon bath. More realistically, phonons
 have spatial structure, with a dispersion $\omega \sim k$. Thus, a TLS at energy $\varepsilon$ couples to a phonon wavepacket correlated over a range $1/\varepsilon$, much larger than the TLS size. The many effective TLSs in the same ``phonon volume'' thus see a spin-phonon coupling  of the form $\hat{X} \sum_i J_i \sigma^x_i \sigma^x_{i+1}$. Integrating out  phonons can thus generate many-spin interactions. Crucially, however,  phonons are non-magnetic and cannot mediate flip-flops among distinct TLSs. Residual magnetic couplings between an effective TLS and its neighbors still involve bond terms 
 $J_i$; provided that $P(J)$ is
 {broad},   relaxation rates {remain} broadly distributed. 
  Hence our conclusions also hold (up to prefactors) for realistic phonon baths.}

Intrinsic energy scales in experimentally-realized Ising magnets such as cobalt niobate~\cite{Coldea177,fava2020glide} lie in the regime accessible to THz tehniques  ($\sim 0.1-10$ THz$\, \sim 5-500$ K). Our results also apply to random Heisenberg antiferromagnets, such as BaCu$_2$Si$_{1-x}$Ge$_x$O$_7$~\cite{PhysRevLett.106.137202} where the doping-dependent exchange scale $\sim400~\text{K}$, and the organic salt quinolinium-(TCNQ)$_2$, where ``random singlet'' physics has been reported~\cite{quinolinium1, quinolinium2} at temperatures $\lesssim 20~\text{K}$.
The remaining  challenge is to drive systems with enough laser power to induce a measurable optical nonlinear magnetic response, a milestone we expect to be reached shortly~\footnote{N. P. Armitage, private communication.}.

\begin{acknowledgments}
We thank N.P.~Armitage, D.~Chaudhuri, F.~Mahmood, R.M.~Nandkishore, E.~Altman, M.~Fava, and V.~Oganesyan for helpful discussions and collaborations on related topics, and N.P. Armitage, M.~Fava, and R.M.~Nandkishore for comments on the manuscript. S.G. acknowledges support from NSF DMR-1653271. S.A.P. acknowledges support from the  European Research Council (ERC) under the European Union Horizon 2020 Research and Innovation Programme [Grant Agreement No.~804213-TMCS], and  EPSRC Grant EP/S020527/1.
\end{acknowledgments}

\bibliography{ising_nl_resp}

\begin{thebibliography}{58}%
\makeatletter
\providecommand \@ifxundefined [1]{%
 \@ifx{#1\undefined}
}%
\providecommand \@ifnum [1]{%
 \ifnum #1\expandafter \@firstoftwo
 \else \expandafter \@secondoftwo
 \fi
}%
\providecommand \@ifx [1]{%
 \ifx #1\expandafter \@firstoftwo
 \else \expandafter \@secondoftwo
 \fi
}%
\providecommand \natexlab [1]{#1}%
\providecommand \enquote  [1]{``#1''}%
\providecommand \bibnamefont  [1]{#1}%
\providecommand \bibfnamefont [1]{#1}%
\providecommand \citenamefont [1]{#1}%
\providecommand \href@noop [0]{\@secondoftwo}%
\providecommand \href [0]{\begingroup \@sanitize@url \@href}%
\providecommand \@href[1]{\@@startlink{#1}\@@href}%
\providecommand \@@href[1]{\endgroup#1\@@endlink}%
\providecommand \@sanitize@url [0]{\catcode `\\12\catcode `\$12\catcode
  `\&12\catcode `\#12\catcode `\^12\catcode `\_12\catcode `\%12\relax}%
\providecommand \@@startlink[1]{}%
\providecommand \@@endlink[0]{}%
\providecommand \url  [0]{\begingroup\@sanitize@url \@url }%
\providecommand \@url [1]{\endgroup\@href {#1}{\urlprefix }}%
\providecommand \urlprefix  [0]{URL }%
\providecommand \Eprint [0]{\href }%
\providecommand \doibase [0]{http://dx.doi.org/}%
\providecommand \selectlanguage [0]{\@gobble}%
\providecommand \bibinfo  [0]{\@secondoftwo}%
\providecommand \bibfield  [0]{\@secondoftwo}%
\providecommand \translation [1]{[#1]}%
\providecommand \BibitemOpen [0]{}%
\providecommand \bibitemStop [0]{}%
\providecommand \bibitemNoStop [0]{.\EOS\space}%
\providecommand \EOS [0]{\spacefactor3000\relax}%
\providecommand \BibitemShut  [1]{\csname bibitem#1\endcsname}%
\let\auto@bib@innerbib\@empty
%</preamble>
\bibitem [{\citenamefont {Cavalleri}\ \emph {et~al.}(2001)\citenamefont
  {Cavalleri}, \citenamefont {T\'oth}, \citenamefont {Siders}, \citenamefont
  {Squier}, \citenamefont {R\'aksi}, \citenamefont {Forget},\ and\
  \citenamefont {Kieffer}}]{PhysRevLett.87.237401}%
  \BibitemOpen
  \bibfield  {author} {\bibinfo {author} {\bibfnamefont {A.}~\bibnamefont
  {Cavalleri}}, \bibinfo {author} {\bibfnamefont {C.}~\bibnamefont {T\'oth}},
  \bibinfo {author} {\bibfnamefont {C.~W.}\ \bibnamefont {Siders}}, \bibinfo
  {author} {\bibfnamefont {J.~A.}\ \bibnamefont {Squier}}, \bibinfo {author}
  {\bibfnamefont {F.}~\bibnamefont {R\'aksi}}, \bibinfo {author} {\bibfnamefont
  {P.}~\bibnamefont {Forget}}, \ and\ \bibinfo {author} {\bibfnamefont {J.~C.}\
  \bibnamefont {Kieffer}},\ }\href {\doibase 10.1103/PhysRevLett.87.237401}
  {\bibfield  {journal} {\bibinfo  {journal} {Phys. Rev. Lett.}\ }\textbf
  {\bibinfo {volume} {87}},\ \bibinfo {pages} {237401} (\bibinfo {year}
  {2001})}\BibitemShut {NoStop}%
\bibitem [{\citenamefont {Ogasawara}\ \emph {et~al.}(2000)\citenamefont
  {Ogasawara}, \citenamefont {Ashida}, \citenamefont {Motoyama}, \citenamefont
  {Eisaki}, \citenamefont {Uchida}, \citenamefont {Tokura}, \citenamefont
  {Ghosh}, \citenamefont {Shukla}, \citenamefont {Mazumdar},\ and\
  \citenamefont {Kuwata-Gonokami}}]{PhysRevLett.85.2204}%
  \BibitemOpen
  \bibfield  {author} {\bibinfo {author} {\bibfnamefont {T.}~\bibnamefont
  {Ogasawara}}, \bibinfo {author} {\bibfnamefont {M.}~\bibnamefont {Ashida}},
  \bibinfo {author} {\bibfnamefont {N.}~\bibnamefont {Motoyama}}, \bibinfo
  {author} {\bibfnamefont {H.}~\bibnamefont {Eisaki}}, \bibinfo {author}
  {\bibfnamefont {S.}~\bibnamefont {Uchida}}, \bibinfo {author} {\bibfnamefont
  {Y.}~\bibnamefont {Tokura}}, \bibinfo {author} {\bibfnamefont
  {H.}~\bibnamefont {Ghosh}}, \bibinfo {author} {\bibfnamefont
  {A.}~\bibnamefont {Shukla}}, \bibinfo {author} {\bibfnamefont
  {S.}~\bibnamefont {Mazumdar}}, \ and\ \bibinfo {author} {\bibfnamefont
  {M.}~\bibnamefont {Kuwata-Gonokami}},\ }\href {\doibase
  10.1103/PhysRevLett.85.2204} {\bibfield  {journal} {\bibinfo  {journal}
  {Phys. Rev. Lett.}\ }\textbf {\bibinfo {volume} {85}},\ \bibinfo {pages}
  {2204} (\bibinfo {year} {2000})}\BibitemShut {NoStop}%
\bibitem [{\citenamefont {Perfetti}\ \emph {et~al.}(2006)\citenamefont
  {Perfetti}, \citenamefont {Loukakos}, \citenamefont {Lisowski}, \citenamefont
  {Bovensiepen}, \citenamefont {Berger}, \citenamefont {Biermann},
  \citenamefont {Cornaglia}, \citenamefont {Georges},\ and\ \citenamefont
  {Wolf}}]{PhysRevLett.97.067402}%
  \BibitemOpen
  \bibfield  {author} {\bibinfo {author} {\bibfnamefont {L.}~\bibnamefont
  {Perfetti}}, \bibinfo {author} {\bibfnamefont {P.~A.}\ \bibnamefont
  {Loukakos}}, \bibinfo {author} {\bibfnamefont {M.}~\bibnamefont {Lisowski}},
  \bibinfo {author} {\bibfnamefont {U.}~\bibnamefont {Bovensiepen}}, \bibinfo
  {author} {\bibfnamefont {H.}~\bibnamefont {Berger}}, \bibinfo {author}
  {\bibfnamefont {S.}~\bibnamefont {Biermann}}, \bibinfo {author}
  {\bibfnamefont {P.~S.}\ \bibnamefont {Cornaglia}}, \bibinfo {author}
  {\bibfnamefont {A.}~\bibnamefont {Georges}}, \ and\ \bibinfo {author}
  {\bibfnamefont {M.}~\bibnamefont {Wolf}},\ }\href {\doibase
  10.1103/PhysRevLett.97.067402} {\bibfield  {journal} {\bibinfo  {journal}
  {Phys. Rev. Lett.}\ }\textbf {\bibinfo {volume} {97}},\ \bibinfo {pages}
  {067402} (\bibinfo {year} {2006})}\BibitemShut {NoStop}%
\bibitem [{\citenamefont {Eckstein}\ and\ \citenamefont
  {Kollar}(2008{\natexlab{a}})}]{PhysRevB.78.245113}%
  \BibitemOpen
  \bibfield  {author} {\bibinfo {author} {\bibfnamefont {M.}~\bibnamefont
  {Eckstein}}\ and\ \bibinfo {author} {\bibfnamefont {M.}~\bibnamefont
  {Kollar}},\ }\href {\doibase 10.1103/PhysRevB.78.245113} {\bibfield
  {journal} {\bibinfo  {journal} {Phys. Rev. B}\ }\textbf {\bibinfo {volume}
  {78}},\ \bibinfo {pages} {245113} (\bibinfo {year}
  {2008}{\natexlab{a}})}\BibitemShut {NoStop}%
\bibitem [{\citenamefont {Eckstein}\ and\ \citenamefont
  {Kollar}(2008{\natexlab{b}})}]{PhysRevB.78.205119}%
  \BibitemOpen
  \bibfield  {author} {\bibinfo {author} {\bibfnamefont {M.}~\bibnamefont
  {Eckstein}}\ and\ \bibinfo {author} {\bibfnamefont {M.}~\bibnamefont
  {Kollar}},\ }\href {\doibase 10.1103/PhysRevB.78.205119} {\bibfield
  {journal} {\bibinfo  {journal} {Phys. Rev. B}\ }\textbf {\bibinfo {volume}
  {78}},\ \bibinfo {pages} {205119} (\bibinfo {year}
  {2008}{\natexlab{b}})}\BibitemShut {NoStop}%
\bibitem [{\citenamefont {Thorsm{\o}lle}\ and\ \citenamefont
  {Armitage}(2010)}]{thorsmolle2010ultrafast}%
  \BibitemOpen
  \bibfield  {author} {\bibinfo {author} {\bibfnamefont {V.}~\bibnamefont
  {Thorsm{\o}lle}}\ and\ \bibinfo {author} {\bibfnamefont {N.}~\bibnamefont
  {Armitage}},\ }\href@noop {} {\bibfield  {journal} {\bibinfo  {journal}
  {Phys. Rev. Lett.}\ }\textbf {\bibinfo {volume} {105}},\ \bibinfo {pages}
  {086601} (\bibinfo {year} {2010})}\BibitemShut {NoStop}%
\bibitem [{\citenamefont {Fausti}\ \emph {et~al.}(2011)\citenamefont {Fausti},
  \citenamefont {Tobey}, \citenamefont {Dean}, \citenamefont {Kaiser},
  \citenamefont {Dienst}, \citenamefont {Hoffmann}, \citenamefont {Pyon},
  \citenamefont {Takayama}, \citenamefont {Takagi},\ and\ \citenamefont
  {Cavalleri}}]{fausti2011light}%
  \BibitemOpen
  \bibfield  {author} {\bibinfo {author} {\bibfnamefont {D.}~\bibnamefont
  {Fausti}}, \bibinfo {author} {\bibfnamefont {R.}~\bibnamefont {Tobey}},
  \bibinfo {author} {\bibfnamefont {N.}~\bibnamefont {Dean}}, \bibinfo {author}
  {\bibfnamefont {S.}~\bibnamefont {Kaiser}}, \bibinfo {author} {\bibfnamefont
  {A.}~\bibnamefont {Dienst}}, \bibinfo {author} {\bibfnamefont {M.~C.}\
  \bibnamefont {Hoffmann}}, \bibinfo {author} {\bibfnamefont {S.}~\bibnamefont
  {Pyon}}, \bibinfo {author} {\bibfnamefont {T.}~\bibnamefont {Takayama}},
  \bibinfo {author} {\bibfnamefont {H.}~\bibnamefont {Takagi}}, \ and\ \bibinfo
  {author} {\bibfnamefont {A.}~\bibnamefont {Cavalleri}},\ }\href@noop {}
  {\bibfield  {journal} {\bibinfo  {journal} {science}\ }\textbf {\bibinfo
  {volume} {331}},\ \bibinfo {pages} {189} (\bibinfo {year}
  {2011})}\BibitemShut {NoStop}%
\bibitem [{\citenamefont {Wang}\ \emph {et~al.}(2013)\citenamefont {Wang},
  \citenamefont {Steinberg}, \citenamefont {Jarillo-Herrero},\ and\
  \citenamefont {Gedik}}]{wang2013observation}%
  \BibitemOpen
  \bibfield  {author} {\bibinfo {author} {\bibfnamefont {Y.}~\bibnamefont
  {Wang}}, \bibinfo {author} {\bibfnamefont {H.}~\bibnamefont {Steinberg}},
  \bibinfo {author} {\bibfnamefont {P.}~\bibnamefont {Jarillo-Herrero}}, \ and\
  \bibinfo {author} {\bibfnamefont {N.}~\bibnamefont {Gedik}},\ }\href@noop {}
  {\bibfield  {journal} {\bibinfo  {journal} {Science}\ }\textbf {\bibinfo
  {volume} {342}},\ \bibinfo {pages} {453} (\bibinfo {year}
  {2013})}\BibitemShut {NoStop}%
\bibitem [{\citenamefont {Mitrano}\ \emph {et~al.}(2016)\citenamefont
  {Mitrano}, \citenamefont {Cantaluppi}, \citenamefont {Nicoletti},
  \citenamefont {Kaiser}, \citenamefont {Perucchi}, \citenamefont {Lupi},
  \citenamefont {Di~Pietro}, \citenamefont {Pontiroli}, \citenamefont
  {Ricc{\`o}}, \citenamefont {Clark} \emph {et~al.}}]{mitrano2016possible}%
  \BibitemOpen
  \bibfield  {author} {\bibinfo {author} {\bibfnamefont {M.}~\bibnamefont
  {Mitrano}}, \bibinfo {author} {\bibfnamefont {A.}~\bibnamefont {Cantaluppi}},
  \bibinfo {author} {\bibfnamefont {D.}~\bibnamefont {Nicoletti}}, \bibinfo
  {author} {\bibfnamefont {S.}~\bibnamefont {Kaiser}}, \bibinfo {author}
  {\bibfnamefont {A.}~\bibnamefont {Perucchi}}, \bibinfo {author}
  {\bibfnamefont {S.}~\bibnamefont {Lupi}}, \bibinfo {author} {\bibfnamefont
  {P.}~\bibnamefont {Di~Pietro}}, \bibinfo {author} {\bibfnamefont
  {D.}~\bibnamefont {Pontiroli}}, \bibinfo {author} {\bibfnamefont
  {M.}~\bibnamefont {Ricc{\`o}}}, \bibinfo {author} {\bibfnamefont {S.~R.}\
  \bibnamefont {Clark}},  \emph {et~al.},\ }\href@noop {} {\bibfield  {journal}
  {\bibinfo  {journal} {Nature}\ }\textbf {\bibinfo {volume} {530}},\ \bibinfo
  {pages} {461} (\bibinfo {year} {2016})}\BibitemShut {NoStop}%
\bibitem [{\citenamefont {Fischer}\ \emph {et~al.}(2016)\citenamefont
  {Fischer}, \citenamefont {Wilson}, \citenamefont {Robles},\ and\
  \citenamefont {Warren}}]{fischer2016invited}%
  \BibitemOpen
  \bibfield  {author} {\bibinfo {author} {\bibfnamefont {M.~C.}\ \bibnamefont
  {Fischer}}, \bibinfo {author} {\bibfnamefont {J.~W.}\ \bibnamefont {Wilson}},
  \bibinfo {author} {\bibfnamefont {F.~E.}\ \bibnamefont {Robles}}, \ and\
  \bibinfo {author} {\bibfnamefont {W.~S.}\ \bibnamefont {Warren}},\
  }\href@noop {} {\bibfield  {journal} {\bibinfo  {journal} {Review of
  Scientific Instruments}\ }\textbf {\bibinfo {volume} {87}},\ \bibinfo {pages}
  {031101} (\bibinfo {year} {2016})}\BibitemShut {NoStop}%
\bibitem [{\citenamefont {Babadi}\ \emph {et~al.}(2017)\citenamefont {Babadi},
  \citenamefont {Knap}, \citenamefont {Martin}, \citenamefont {Refael},\ and\
  \citenamefont {Demler}}]{babadi1}%
  \BibitemOpen
  \bibfield  {author} {\bibinfo {author} {\bibfnamefont {M.}~\bibnamefont
  {Babadi}}, \bibinfo {author} {\bibfnamefont {M.}~\bibnamefont {Knap}},
  \bibinfo {author} {\bibfnamefont {I.}~\bibnamefont {Martin}}, \bibinfo
  {author} {\bibfnamefont {G.}~\bibnamefont {Refael}}, \ and\ \bibinfo {author}
  {\bibfnamefont {E.}~\bibnamefont {Demler}},\ }\href {\doibase
  10.1103/PhysRevB.96.014512} {\bibfield  {journal} {\bibinfo  {journal} {Phys.
  Rev. B}\ }\textbf {\bibinfo {volume} {96}},\ \bibinfo {pages} {014512}
  (\bibinfo {year} {2017})}\BibitemShut {NoStop}%
\bibitem [{\citenamefont {Mukamel}(1995)}]{Mukamel1995}%
  \BibitemOpen
  \bibfield  {author} {\bibinfo {author} {\bibfnamefont {S.}~\bibnamefont
  {Mukamel}},\ }\href@noop {} {\emph {\bibinfo {title} {Principles of
  {N}onlinear {O}ptical {S}pectroscopy}}}\ (\bibinfo  {publisher} {Oxford
  University Press},\ \bibinfo {address} {New York},\ \bibinfo {year}
  {1995})\BibitemShut {NoStop}%
\bibitem [{\citenamefont {Hamm}(2005)}]{hamm2005principles}%
  \BibitemOpen
  \bibfield  {author} {\bibinfo {author} {\bibfnamefont {P.}~\bibnamefont
  {Hamm}},\ }\href {http://www.mitr.p.lodz.pl/evu/lectures/Hamm.pdf} {\enquote
  {\bibinfo {title} {Principles of nonlinear optical spectroscopy: A practical
  approach or: Mukamel for dummies},}\ } (\bibinfo {year} {2005})\BibitemShut
  {NoStop}%
\bibitem [{\citenamefont {Axt}\ and\ \citenamefont
  {Kuhn}(2004)}]{axt2004femtosecond}%
  \BibitemOpen
  \bibfield  {author} {\bibinfo {author} {\bibfnamefont {V.~M.}\ \bibnamefont
  {Axt}}\ and\ \bibinfo {author} {\bibfnamefont {T.}~\bibnamefont {Kuhn}},\
  }\href@noop {} {\bibfield  {journal} {\bibinfo  {journal} {Reports on
  Progress in Physics}\ }\textbf {\bibinfo {volume} {67}},\ \bibinfo {pages}
  {433} (\bibinfo {year} {2004})}\BibitemShut {NoStop}%
\bibitem [{\citenamefont {Cundiff}\ and\ \citenamefont
  {Mukamel}(2013)}]{cundiff2013optical}%
  \BibitemOpen
  \bibfield  {author} {\bibinfo {author} {\bibfnamefont {S.~T.}\ \bibnamefont
  {Cundiff}}\ and\ \bibinfo {author} {\bibfnamefont {S.}~\bibnamefont
  {Mukamel}},\ }\href {\doibase 10.1063/PT.3.2047} {\bibfield  {journal}
  {\bibinfo  {journal} {Physics Today}\ }\textbf {\bibinfo {volume} {66}},\
  \bibinfo {pages} {44} (\bibinfo {year} {2013})}\BibitemShut {NoStop}%
\bibitem [{\citenamefont {Woerner}\ \emph {et~al.}(2013)\citenamefont
  {Woerner}, \citenamefont {Kuehn}, \citenamefont {Bowlan}, \citenamefont
  {Reimann},\ and\ \citenamefont {Elsaesser}}]{woerner2013ultrafast}%
  \BibitemOpen
  \bibfield  {author} {\bibinfo {author} {\bibfnamefont {M.}~\bibnamefont
  {Woerner}}, \bibinfo {author} {\bibfnamefont {W.}~\bibnamefont {Kuehn}},
  \bibinfo {author} {\bibfnamefont {P.}~\bibnamefont {Bowlan}}, \bibinfo
  {author} {\bibfnamefont {K.}~\bibnamefont {Reimann}}, \ and\ \bibinfo
  {author} {\bibfnamefont {T.}~\bibnamefont {Elsaesser}},\ }\href@noop {}
  {\bibfield  {journal} {\bibinfo  {journal} {New J. Phys.}\ }\textbf {\bibinfo
  {volume} {15}},\ \bibinfo {pages} {025039} (\bibinfo {year}
  {2013})}\BibitemShut {NoStop}%
\bibitem [{\citenamefont {Faoro}\ and\ \citenamefont
  {Ioffe}(2012)}]{PhysRevLett.109.157005}%
  \BibitemOpen
  \bibfield  {author} {\bibinfo {author} {\bibfnamefont {L.}~\bibnamefont
  {Faoro}}\ and\ \bibinfo {author} {\bibfnamefont {L.~B.}\ \bibnamefont
  {Ioffe}},\ }\href {\doibase 10.1103/PhysRevLett.109.157005} {\bibfield
  {journal} {\bibinfo  {journal} {Phys. Rev. Lett.}\ }\textbf {\bibinfo
  {volume} {109}},\ \bibinfo {pages} {157005} (\bibinfo {year}
  {2012})}\BibitemShut {NoStop}%
\bibitem [{\citenamefont {Rosenow}\ and\ \citenamefont
  {Nattermann}(2006)}]{rosenow}%
  \BibitemOpen
  \bibfield  {author} {\bibinfo {author} {\bibfnamefont {B.}~\bibnamefont
  {Rosenow}}\ and\ \bibinfo {author} {\bibfnamefont {T.}~\bibnamefont
  {Nattermann}},\ }\href {\doibase 10.1103/PhysRevB.73.085103} {\bibfield
  {journal} {\bibinfo  {journal} {Phys. Rev. B}\ }\textbf {\bibinfo {volume}
  {73}},\ \bibinfo {pages} {085103} (\bibinfo {year} {2006})}\BibitemShut
  {NoStop}%
\bibitem [{\citenamefont {Gopalakrishnan}\ \emph {et~al.}(2016)\citenamefont
  {Gopalakrishnan}, \citenamefont {Knap},\ and\ \citenamefont
  {Demler}}]{sg_heating}%
  \BibitemOpen
  \bibfield  {author} {\bibinfo {author} {\bibfnamefont {S.}~\bibnamefont
  {Gopalakrishnan}}, \bibinfo {author} {\bibfnamefont {M.}~\bibnamefont
  {Knap}}, \ and\ \bibinfo {author} {\bibfnamefont {E.}~\bibnamefont
  {Demler}},\ }\href {\doibase 10.1103/PhysRevB.94.094201} {\bibfield
  {journal} {\bibinfo  {journal} {Phys. Rev. B}\ }\textbf {\bibinfo {volume}
  {94}},\ \bibinfo {pages} {094201} (\bibinfo {year} {2016})}\BibitemShut
  {NoStop}%
\bibitem [{\citenamefont {Kozarzewski}\ \emph {et~al.}(2016)\citenamefont
  {Kozarzewski}, \citenamefont {Prelov\ifmmode~\check{s}\else \v{s}\fi{}ek},\
  and\ \citenamefont {Mierzejewski}}]{kozarzewski}%
  \BibitemOpen
  \bibfield  {author} {\bibinfo {author} {\bibfnamefont {M.}~\bibnamefont
  {Kozarzewski}}, \bibinfo {author} {\bibfnamefont {P.}~\bibnamefont
  {Prelov\ifmmode~\check{s}\else \v{s}\fi{}ek}}, \ and\ \bibinfo {author}
  {\bibfnamefont {M.}~\bibnamefont {Mierzejewski}},\ }\href {\doibase
  10.1103/PhysRevB.93.235151} {\bibfield  {journal} {\bibinfo  {journal} {Phys.
  Rev. B}\ }\textbf {\bibinfo {volume} {93}},\ \bibinfo {pages} {235151}
  (\bibinfo {year} {2016})}\BibitemShut {NoStop}%
\bibitem [{\citenamefont {Rehn}\ \emph {et~al.}(2016)\citenamefont {Rehn},
  \citenamefont {Lazarides}, \citenamefont {Pollmann},\ and\ \citenamefont
  {Moessner}}]{rehn2016}%
  \BibitemOpen
  \bibfield  {author} {\bibinfo {author} {\bibfnamefont {J.}~\bibnamefont
  {Rehn}}, \bibinfo {author} {\bibfnamefont {A.}~\bibnamefont {Lazarides}},
  \bibinfo {author} {\bibfnamefont {F.}~\bibnamefont {Pollmann}}, \ and\
  \bibinfo {author} {\bibfnamefont {R.}~\bibnamefont {Moessner}},\ }\href
  {\doibase 10.1103/PhysRevB.94.020201} {\bibfield  {journal} {\bibinfo
  {journal} {Phys. Rev. B}\ }\textbf {\bibinfo {volume} {94}},\ \bibinfo
  {pages} {020201} (\bibinfo {year} {2016})}\BibitemShut {NoStop}%
\bibitem [{\citenamefont {Liu}\ \emph {et~al.}(2018)\citenamefont {Liu},
  \citenamefont {Chalker}, \citenamefont {Khemani},\ and\ \citenamefont
  {Sondhi}}]{lcks}%
  \BibitemOpen
  \bibfield  {author} {\bibinfo {author} {\bibfnamefont {D.~T.}\ \bibnamefont
  {Liu}}, \bibinfo {author} {\bibfnamefont {J.~T.}\ \bibnamefont {Chalker}},
  \bibinfo {author} {\bibfnamefont {V.}~\bibnamefont {Khemani}}, \ and\
  \bibinfo {author} {\bibfnamefont {S.~L.}\ \bibnamefont {Sondhi}},\ }\href
  {\doibase 10.1103/PhysRevB.98.214202} {\bibfield  {journal} {\bibinfo
  {journal} {Phys. Rev. B}\ }\textbf {\bibinfo {volume} {98}},\ \bibinfo
  {pages} {214202} (\bibinfo {year} {2018})}\BibitemShut {NoStop}%
\bibitem [{\citenamefont {Chan}\ \emph {et~al.}(2019)\citenamefont {Chan},
  \citenamefont {De~Luca},\ and\ \citenamefont
  {Chalker}}]{PhysRevLett.122.220601}%
  \BibitemOpen
  \bibfield  {author} {\bibinfo {author} {\bibfnamefont {A.}~\bibnamefont
  {Chan}}, \bibinfo {author} {\bibfnamefont {A.}~\bibnamefont {De~Luca}}, \
  and\ \bibinfo {author} {\bibfnamefont {J.~T.}\ \bibnamefont {Chalker}},\
  }\href {\doibase 10.1103/PhysRevLett.122.220601} {\bibfield  {journal}
  {\bibinfo  {journal} {Phys. Rev. Lett.}\ }\textbf {\bibinfo {volume} {122}},\
  \bibinfo {pages} {220601} (\bibinfo {year} {2019})}\BibitemShut {NoStop}%
\bibitem [{\citenamefont {Burin}\ and\ \citenamefont
  {Maksymov}(2018)}]{PhysRevB.97.214208}%
  \BibitemOpen
  \bibfield  {author} {\bibinfo {author} {\bibfnamefont {A.~L.}\ \bibnamefont
  {Burin}}\ and\ \bibinfo {author} {\bibfnamefont {A.~O.}\ \bibnamefont
  {Maksymov}},\ }\href {\doibase 10.1103/PhysRevB.97.214208} {\bibfield
  {journal} {\bibinfo  {journal} {Phys. Rev. B}\ }\textbf {\bibinfo {volume}
  {97}},\ \bibinfo {pages} {214208} (\bibinfo {year} {2018})}\BibitemShut
  {NoStop}%
\bibitem [{\citenamefont {Wan}\ and\ \citenamefont
  {Armitage}(2019)}]{wan2019resolving}%
  \BibitemOpen
  \bibfield  {author} {\bibinfo {author} {\bibfnamefont {Y.}~\bibnamefont
  {Wan}}\ and\ \bibinfo {author} {\bibfnamefont {N.}~\bibnamefont {Armitage}},\
  }\href@noop {} {\bibfield  {journal} {\bibinfo  {journal} {Phys. Rev. Lett.}\
  }\textbf {\bibinfo {volume} {122}},\ \bibinfo {pages} {257401} (\bibinfo
  {year} {2019})}\BibitemShut {NoStop}%
\bibitem [{\citenamefont {Choi}\ \emph {et~al.}(2020)\citenamefont {Choi},
  \citenamefont {Lee},\ and\ \citenamefont {Kim}}]{choi2020theory}%
  \BibitemOpen
  \bibfield  {author} {\bibinfo {author} {\bibfnamefont {W.}~\bibnamefont
  {Choi}}, \bibinfo {author} {\bibfnamefont {K.~H.}\ \bibnamefont {Lee}}, \
  and\ \bibinfo {author} {\bibfnamefont {Y.~B.}\ \bibnamefont {Kim}},\
  }\href@noop {} {\bibfield  {journal} {\bibinfo  {journal} {Phys. Rev. Lett.}\
  }\textbf {\bibinfo {volume} {124}},\ \bibinfo {pages} {117205} (\bibinfo
  {year} {2020})}\BibitemShut {NoStop}%
\bibitem [{\citenamefont {Mahmood}\ \emph {et~al.}(2020)\citenamefont
  {Mahmood}, \citenamefont {Chaudhuri}, \citenamefont {Gopalakrishnan},
  \citenamefont {Nandkishore},\ and\ \citenamefont
  {Armitage}}]{mahmood2020observation}%
  \BibitemOpen
  \bibfield  {author} {\bibinfo {author} {\bibfnamefont {F.}~\bibnamefont
  {Mahmood}}, \bibinfo {author} {\bibfnamefont {D.}~\bibnamefont {Chaudhuri}},
  \bibinfo {author} {\bibfnamefont {S.}~\bibnamefont {Gopalakrishnan}},
  \bibinfo {author} {\bibfnamefont {R.}~\bibnamefont {Nandkishore}}, \ and\
  \bibinfo {author} {\bibfnamefont {N.}~\bibnamefont {Armitage}},\ }\href@noop
  {} {\bibfield  {journal} {\bibinfo  {journal} {arXiv preprint
  arXiv:2005.10822}\ } (\bibinfo {year} {2020})}\BibitemShut {NoStop}%
\bibitem [{\citenamefont {M{\"u}ller}\ \emph {et~al.}(2019)\citenamefont
  {M{\"u}ller}, \citenamefont {Cole},\ and\ \citenamefont
  {Lisenfeld}}]{muller2019towards}%
  \BibitemOpen
  \bibfield  {author} {\bibinfo {author} {\bibfnamefont {C.}~\bibnamefont
  {M{\"u}ller}}, \bibinfo {author} {\bibfnamefont {J.~H.}\ \bibnamefont
  {Cole}}, \ and\ \bibinfo {author} {\bibfnamefont {J.}~\bibnamefont
  {Lisenfeld}},\ }\href@noop {} {\bibfield  {journal} {\bibinfo  {journal}
  {Reports on Progress in Physics}\ }\textbf {\bibinfo {volume} {82}},\
  \bibinfo {pages} {124501} (\bibinfo {year} {2019})}\BibitemShut {NoStop}%
\bibitem [{\citenamefont {Ma}\ \emph {et~al.}(1979)\citenamefont {Ma},
  \citenamefont {Dasgupta},\ and\ \citenamefont {Hu}}]{ma1979random}%
  \BibitemOpen
  \bibfield  {author} {\bibinfo {author} {\bibfnamefont {S.-k.}\ \bibnamefont
  {Ma}}, \bibinfo {author} {\bibfnamefont {C.}~\bibnamefont {Dasgupta}}, \ and\
  \bibinfo {author} {\bibfnamefont {C.-k.}\ \bibnamefont {Hu}},\ }\href@noop {}
  {\bibfield  {journal} {\bibinfo  {journal} {Physical review letters}\
  }\textbf {\bibinfo {volume} {43}},\ \bibinfo {pages} {1434} (\bibinfo {year}
  {1979})}\BibitemShut {NoStop}%
\bibitem [{\citenamefont {Fisher}(1992)}]{fisher1992random}%
  \BibitemOpen
  \bibfield  {author} {\bibinfo {author} {\bibfnamefont {D.~S.}\ \bibnamefont
  {Fisher}},\ }\href@noop {} {\bibfield  {journal} {\bibinfo  {journal}
  {Physical review letters}\ }\textbf {\bibinfo {volume} {69}},\ \bibinfo
  {pages} {534} (\bibinfo {year} {1992})}\BibitemShut {NoStop}%
\bibitem [{\citenamefont {Fisher}(1995)}]{fisher1995critical}%
  \BibitemOpen
  \bibfield  {author} {\bibinfo {author} {\bibfnamefont {D.~S.}\ \bibnamefont
  {Fisher}},\ }\href@noop {} {\bibfield  {journal} {\bibinfo  {journal}
  {Physical review b}\ }\textbf {\bibinfo {volume} {51}},\ \bibinfo {pages}
  {6411} (\bibinfo {year} {1995})}\BibitemShut {NoStop}%
\bibitem [{\citenamefont {Motrunich}\ \emph {et~al.}(2000)\citenamefont
  {Motrunich}, \citenamefont {Mau}, \citenamefont {Huse},\ and\ \citenamefont
  {Fisher}}]{motrunich2000infinite}%
  \BibitemOpen
  \bibfield  {author} {\bibinfo {author} {\bibfnamefont {O.}~\bibnamefont
  {Motrunich}}, \bibinfo {author} {\bibfnamefont {S.-C.}\ \bibnamefont {Mau}},
  \bibinfo {author} {\bibfnamefont {D.~A.}\ \bibnamefont {Huse}}, \ and\
  \bibinfo {author} {\bibfnamefont {D.~S.}\ \bibnamefont {Fisher}},\
  }\href@noop {} {\bibfield  {journal} {\bibinfo  {journal} {Physical Review
  B}\ }\textbf {\bibinfo {volume} {61}},\ \bibinfo {pages} {1160} (\bibinfo
  {year} {2000})}\BibitemShut {NoStop}%
\bibitem [{\citenamefont {Motrunich}\ \emph {et~al.}(2001)\citenamefont
  {Motrunich}, \citenamefont {Damle},\ and\ \citenamefont {Huse}}]{mdh}%
  \BibitemOpen
  \bibfield  {author} {\bibinfo {author} {\bibfnamefont {O.}~\bibnamefont
  {Motrunich}}, \bibinfo {author} {\bibfnamefont {K.}~\bibnamefont {Damle}}, \
  and\ \bibinfo {author} {\bibfnamefont {D.~A.}\ \bibnamefont {Huse}},\ }\href
  {\doibase 10.1103/PhysRevB.63.134424} {\bibfield  {journal} {\bibinfo
  {journal} {Phys. Rev. B}\ }\textbf {\bibinfo {volume} {63}},\ \bibinfo
  {pages} {134424} (\bibinfo {year} {2001})}\BibitemShut {NoStop}%
\bibitem [{\citenamefont {Shiroka}\ \emph {et~al.}(2011)\citenamefont
  {Shiroka}, \citenamefont {Casola}, \citenamefont {Glazkov}, \citenamefont
  {Zheludev}, \citenamefont {Pr\ifmmode~\check{s}\else \v{s}\fi{}a},
  \citenamefont {Ott},\ and\ \citenamefont {Mesot}}]{PhysRevLett.106.137202}%
  \BibitemOpen
  \bibfield  {author} {\bibinfo {author} {\bibfnamefont {T.}~\bibnamefont
  {Shiroka}}, \bibinfo {author} {\bibfnamefont {F.}~\bibnamefont {Casola}},
  \bibinfo {author} {\bibfnamefont {V.}~\bibnamefont {Glazkov}}, \bibinfo
  {author} {\bibfnamefont {A.}~\bibnamefont {Zheludev}}, \bibinfo {author}
  {\bibfnamefont {K.}~\bibnamefont {Pr\ifmmode~\check{s}\else \v{s}\fi{}a}},
  \bibinfo {author} {\bibfnamefont {H.-R.}\ \bibnamefont {Ott}}, \ and\
  \bibinfo {author} {\bibfnamefont {J.}~\bibnamefont {Mesot}},\ }\href
  {\doibase 10.1103/PhysRevLett.106.137202} {\bibfield  {journal} {\bibinfo
  {journal} {Phys. Rev. Lett.}\ }\textbf {\bibinfo {volume} {106}},\ \bibinfo
  {pages} {137202} (\bibinfo {year} {2011})}\BibitemShut {NoStop}%
\bibitem [{\citenamefont {Herbrych}\ \emph {et~al.}(2013)\citenamefont
  {Herbrych}, \citenamefont {Kokalj},\ and\ \citenamefont
  {Prelov\ifmmode~\check{s}\else \v{s}\fi{}ek}}]{PhysRevLett.111.147203}%
  \BibitemOpen
  \bibfield  {author} {\bibinfo {author} {\bibfnamefont {J.}~\bibnamefont
  {Herbrych}}, \bibinfo {author} {\bibfnamefont {J.}~\bibnamefont {Kokalj}}, \
  and\ \bibinfo {author} {\bibfnamefont {P.}~\bibnamefont
  {Prelov\ifmmode~\check{s}\else \v{s}\fi{}ek}},\ }\href {\doibase
  10.1103/PhysRevLett.111.147203} {\bibfield  {journal} {\bibinfo  {journal}
  {Phys. Rev. Lett.}\ }\textbf {\bibinfo {volume} {111}},\ \bibinfo {pages}
  {147203} (\bibinfo {year} {2013})}\BibitemShut {NoStop}%
\bibitem [{\citenamefont {Fisher}(1999)}]{fisher1999phase}%
  \BibitemOpen
  \bibfield  {author} {\bibinfo {author} {\bibfnamefont {D.~S.}\ \bibnamefont
  {Fisher}},\ }\href@noop {} {\bibfield  {journal} {\bibinfo  {journal}
  {Physica A: Statistical Mechanics and its Applications}\ }\textbf {\bibinfo
  {volume} {263}},\ \bibinfo {pages} {222} (\bibinfo {year}
  {1999})}\BibitemShut {NoStop}%
\bibitem [{\citenamefont {Basko}\ \emph {et~al.}(2006)\citenamefont {Basko},
  \citenamefont {Aleiner},\ and\ \citenamefont {Altshuler}}]{basko2006metal}%
  \BibitemOpen
  \bibfield  {author} {\bibinfo {author} {\bibfnamefont {D.~M.}\ \bibnamefont
  {Basko}}, \bibinfo {author} {\bibfnamefont {I.~L.}\ \bibnamefont {Aleiner}},
  \ and\ \bibinfo {author} {\bibfnamefont {B.~L.}\ \bibnamefont {Altshuler}},\
  }\href@noop {} {\bibfield  {journal} {\bibinfo  {journal} {Ann. Physics}\
  }\textbf {\bibinfo {volume} {321}},\ \bibinfo {pages} {1126} (\bibinfo {year}
  {2006})}\BibitemShut {NoStop}%
\bibitem [{\citenamefont {Bar~Lev}\ and\ \citenamefont
  {Reichman}(2014)}]{PhysRevB.89.220201}%
  \BibitemOpen
  \bibfield  {author} {\bibinfo {author} {\bibfnamefont {Y.}~\bibnamefont
  {Bar~Lev}}\ and\ \bibinfo {author} {\bibfnamefont {D.~R.}\ \bibnamefont
  {Reichman}},\ }\href {\doibase 10.1103/PhysRevB.89.220201} {\bibfield
  {journal} {\bibinfo  {journal} {Phys. Rev. B}\ }\textbf {\bibinfo {volume}
  {89}},\ \bibinfo {pages} {220201} (\bibinfo {year} {2014})}\BibitemShut
  {NoStop}%
\bibitem [{\citenamefont {Gopalakrishnan}\ and\ \citenamefont
  {Nandkishore}(2014)}]{gn}%
  \BibitemOpen
  \bibfield  {author} {\bibinfo {author} {\bibfnamefont {S.}~\bibnamefont
  {Gopalakrishnan}}\ and\ \bibinfo {author} {\bibfnamefont {R.}~\bibnamefont
  {Nandkishore}},\ }\href {\doibase 10.1103/PhysRevB.90.224203} {\bibfield
  {journal} {\bibinfo  {journal} {Phys. Rev. B}\ }\textbf {\bibinfo {volume}
  {90}},\ \bibinfo {pages} {224203} (\bibinfo {year} {2014})}\BibitemShut
  {NoStop}%
\bibitem [{\citenamefont {Leggett}\ \emph {et~al.}(1987)\citenamefont
  {Leggett}, \citenamefont {Chakravarty}, \citenamefont {Dorsey}, \citenamefont
  {Fisher}, \citenamefont {Garg},\ and\ \citenamefont
  {Zwerger}}]{leggett_review}%
  \BibitemOpen
  \bibfield  {author} {\bibinfo {author} {\bibfnamefont {A.~J.}\ \bibnamefont
  {Leggett}}, \bibinfo {author} {\bibfnamefont {S.}~\bibnamefont
  {Chakravarty}}, \bibinfo {author} {\bibfnamefont {A.~T.}\ \bibnamefont
  {Dorsey}}, \bibinfo {author} {\bibfnamefont {M.~P.~A.}\ \bibnamefont
  {Fisher}}, \bibinfo {author} {\bibfnamefont {A.}~\bibnamefont {Garg}}, \ and\
  \bibinfo {author} {\bibfnamefont {W.}~\bibnamefont {Zwerger}},\ }\href
  {\doibase 10.1103/RevModPhys.59.1} {\bibfield  {journal} {\bibinfo  {journal}
  {Rev. Mod. Phys.}\ }\textbf {\bibinfo {volume} {59}},\ \bibinfo {pages} {1}
  (\bibinfo {year} {1987})}\BibitemShut {NoStop}%
\bibitem [{\citenamefont {Millis}\ \emph {et~al.}(2001)\citenamefont {Millis},
  \citenamefont {Morr},\ and\ \citenamefont {Schmalian}}]{mms1}%
  \BibitemOpen
  \bibfield  {author} {\bibinfo {author} {\bibfnamefont {A.~J.}\ \bibnamefont
  {Millis}}, \bibinfo {author} {\bibfnamefont {D.~K.}\ \bibnamefont {Morr}}, \
  and\ \bibinfo {author} {\bibfnamefont {J.}~\bibnamefont {Schmalian}},\ }\href
  {\doibase 10.1103/PhysRevLett.87.167202} {\bibfield  {journal} {\bibinfo
  {journal} {Phys. Rev. Lett.}\ }\textbf {\bibinfo {volume} {87}},\ \bibinfo
  {pages} {167202} (\bibinfo {year} {2001})}\BibitemShut {NoStop}%
\bibitem [{\citenamefont {Millis}\ \emph {et~al.}(2002)\citenamefont {Millis},
  \citenamefont {Morr},\ and\ \citenamefont {Schmalian}}]{mms2}%
  \BibitemOpen
  \bibfield  {author} {\bibinfo {author} {\bibfnamefont {A.~J.}\ \bibnamefont
  {Millis}}, \bibinfo {author} {\bibfnamefont {D.~K.}\ \bibnamefont {Morr}}, \
  and\ \bibinfo {author} {\bibfnamefont {J.}~\bibnamefont {Schmalian}},\ }\href
  {\doibase 10.1103/PhysRevB.66.174433} {\bibfield  {journal} {\bibinfo
  {journal} {Phys. Rev. B}\ }\textbf {\bibinfo {volume} {66}},\ \bibinfo
  {pages} {174433} (\bibinfo {year} {2002})}\BibitemShut {NoStop}%
\bibitem [{\citenamefont {Vojta}(2003)}]{vojta2003}%
  \BibitemOpen
  \bibfield  {author} {\bibinfo {author} {\bibfnamefont {T.}~\bibnamefont
  {Vojta}},\ }\href {\doibase 10.1103/PhysRevLett.90.107202} {\bibfield
  {journal} {\bibinfo  {journal} {Phys. Rev. Lett.}\ }\textbf {\bibinfo
  {volume} {90}},\ \bibinfo {pages} {107202} (\bibinfo {year}
  {2003})}\BibitemShut {NoStop}%
\bibitem [{\citenamefont {Schehr}\ and\ \citenamefont
  {Rieger}(2006)}]{schehr2006}%
  \BibitemOpen
  \bibfield  {author} {\bibinfo {author} {\bibfnamefont {G.}~\bibnamefont
  {Schehr}}\ and\ \bibinfo {author} {\bibfnamefont {H.}~\bibnamefont
  {Rieger}},\ }\href {\doibase 10.1103/PhysRevLett.96.227201} {\bibfield
  {journal} {\bibinfo  {journal} {Phys. Rev. Lett.}\ }\textbf {\bibinfo
  {volume} {96}},\ \bibinfo {pages} {227201} (\bibinfo {year}
  {2006})}\BibitemShut {NoStop}%
\bibitem [{\citenamefont {Hoyos}\ and\ \citenamefont
  {Vojta}(2008)}]{hoyos2008}%
  \BibitemOpen
  \bibfield  {author} {\bibinfo {author} {\bibfnamefont {J.~A.}\ \bibnamefont
  {Hoyos}}\ and\ \bibinfo {author} {\bibfnamefont {T.}~\bibnamefont {Vojta}},\
  }\href {\doibase 10.1103/PhysRevLett.100.240601} {\bibfield  {journal}
  {\bibinfo  {journal} {Phys. Rev. Lett.}\ }\textbf {\bibinfo {volume} {100}},\
  \bibinfo {pages} {240601} (\bibinfo {year} {2008})}\BibitemShut {NoStop}%
\bibitem [{\citenamefont {Hoyos}\ and\ \citenamefont
  {Vojta}(2012)}]{hoyos2012}%
  \BibitemOpen
  \bibfield  {author} {\bibinfo {author} {\bibfnamefont {J.~A.}\ \bibnamefont
  {Hoyos}}\ and\ \bibinfo {author} {\bibfnamefont {T.}~\bibnamefont {Vojta}},\
  }\href {\doibase 10.1103/PhysRevB.85.174403} {\bibfield  {journal} {\bibinfo
  {journal} {Phys. Rev. B}\ }\textbf {\bibinfo {volume} {85}},\ \bibinfo
  {pages} {174403} (\bibinfo {year} {2012})}\BibitemShut {NoStop}%
\bibitem [{\citenamefont {Chin}\ \emph {et~al.}(2012)\citenamefont {Chin},
  \citenamefont {Huelga},\ and\ \citenamefont
  {Plenio}}]{PhysRevLett.109.233601}%
  \BibitemOpen
  \bibfield  {author} {\bibinfo {author} {\bibfnamefont {A.~W.}\ \bibnamefont
  {Chin}}, \bibinfo {author} {\bibfnamefont {S.~F.}\ \bibnamefont {Huelga}}, \
  and\ \bibinfo {author} {\bibfnamefont {M.~B.}\ \bibnamefont {Plenio}},\
  }\href {\doibase 10.1103/PhysRevLett.109.233601} {\bibfield  {journal}
  {\bibinfo  {journal} {Phys. Rev. Lett.}\ }\textbf {\bibinfo {volume} {109}},\
  \bibinfo {pages} {233601} (\bibinfo {year} {2012})}\BibitemShut {NoStop}%
\bibitem [{\citenamefont {\ifmmode~\check{S}\else \v{S}\fi{}anda}\ and\
  \citenamefont {Mukamel}(2007)}]{PhysRevLett.98.080603}%
  \BibitemOpen
  \bibfield  {author} {\bibinfo {author} {\bibfnamefont {F.~c.~v.}\
  \bibnamefont {\ifmmode~\check{S}\else \v{S}\fi{}anda}}\ and\ \bibinfo
  {author} {\bibfnamefont {S.}~\bibnamefont {Mukamel}},\ }\href {\doibase
  10.1103/PhysRevLett.98.080603} {\bibfield  {journal} {\bibinfo  {journal}
  {Phys. Rev. Lett.}\ }\textbf {\bibinfo {volume} {98}},\ \bibinfo {pages}
  {080603} (\bibinfo {year} {2007})}\BibitemShut {NoStop}%
\bibitem [{\citenamefont {Fisher}(1994)}]{XXZ_Fisher_RG}%
  \BibitemOpen
  \bibfield  {author} {\bibinfo {author} {\bibfnamefont {D.~S.}\ \bibnamefont
  {Fisher}},\ }\href {\doibase 10.1103/PhysRevB.50.3799} {\bibfield  {journal}
  {\bibinfo  {journal} {Phys. Rev. B}\ }\textbf {\bibinfo {volume} {50}},\
  \bibinfo {pages} {3799} (\bibinfo {year} {1994})}\BibitemShut {NoStop}%
\bibitem [{\citenamefont {Senthil}\ and\ \citenamefont
  {Majumdar}(1996)}]{PhysRevLett.76.3001}%
  \BibitemOpen
  \bibfield  {author} {\bibinfo {author} {\bibfnamefont {T.}~\bibnamefont
  {Senthil}}\ and\ \bibinfo {author} {\bibfnamefont {S.~N.}\ \bibnamefont
  {Majumdar}},\ }\href {\doibase 10.1103/PhysRevLett.76.3001} {\bibfield
  {journal} {\bibinfo  {journal} {Phys. Rev. Lett.}\ }\textbf {\bibinfo
  {volume} {76}},\ \bibinfo {pages} {3001} (\bibinfo {year}
  {1996})}\BibitemShut {NoStop}%
\bibitem [{\citenamefont {Hyman}\ and\ \citenamefont
  {Yang}(1997)}]{hyman_yang}%
  \BibitemOpen
  \bibfield  {author} {\bibinfo {author} {\bibfnamefont {R.~A.}\ \bibnamefont
  {Hyman}}\ and\ \bibinfo {author} {\bibfnamefont {K.}~\bibnamefont {Yang}},\
  }\href {\doibase 10.1103/PhysRevLett.78.1783} {\bibfield  {journal} {\bibinfo
   {journal} {Phys. Rev. Lett.}\ }\textbf {\bibinfo {volume} {78}},\ \bibinfo
  {pages} {1783} (\bibinfo {year} {1997})}\BibitemShut {NoStop}%
\bibitem [{\citenamefont {{Kang}}\ \emph {et~al.}(2020)\citenamefont {{Kang}},
  \citenamefont {{Parameswaran}}, \citenamefont {{Potter}}, \citenamefont
  {{Vasseur}},\ and\ \citenamefont {{Gazit}}}]{2020arXiv200809617K}%
  \BibitemOpen
  \bibfield  {author} {\bibinfo {author} {\bibfnamefont {B.}~\bibnamefont
  {{Kang}}}, \bibinfo {author} {\bibfnamefont {S.~A.}\ \bibnamefont
  {{Parameswaran}}}, \bibinfo {author} {\bibfnamefont {A.~C.}\ \bibnamefont
  {{Potter}}}, \bibinfo {author} {\bibfnamefont {R.}~\bibnamefont {{Vasseur}}},
  \ and\ \bibinfo {author} {\bibfnamefont {S.}~\bibnamefont {{Gazit}}},\
  }\href@noop {} {\bibfield  {journal} {\bibinfo  {journal} {arXiv e-prints}\
  ,\ \bibinfo {eid} {arXiv:2008.09617}} (\bibinfo {year} {2020})},\ \Eprint
  {http://arxiv.org/abs/2008.09617} {arXiv:2008.09617 [cond-mat.str-el]}
  \BibitemShut {NoStop}%
\bibitem [{\citenamefont {Bhatt}\ and\ \citenamefont {Lee}(1982)}]{bhattlee}%
  \BibitemOpen
  \bibfield  {author} {\bibinfo {author} {\bibfnamefont {R.~N.}\ \bibnamefont
  {Bhatt}}\ and\ \bibinfo {author} {\bibfnamefont {P.~A.}\ \bibnamefont
  {Lee}},\ }\href {\doibase 10.1103/PhysRevLett.48.344} {\bibfield  {journal}
  {\bibinfo  {journal} {Phys. Rev. Lett.}\ }\textbf {\bibinfo {volume} {48}},\
  \bibinfo {pages} {344} (\bibinfo {year} {1982})}\BibitemShut {NoStop}%
\bibitem [{\citenamefont {Coldea}\ \emph {et~al.}(2010)\citenamefont {Coldea},
  \citenamefont {Tennant}, \citenamefont {Wheeler}, \citenamefont {Wawrzynska},
  \citenamefont {Prabhakaran}, \citenamefont {Telling}, \citenamefont
  {Habicht}, \citenamefont {Smeibidl},\ and\ \citenamefont
  {Kiefer}}]{Coldea177}%
  \BibitemOpen
  \bibfield  {author} {\bibinfo {author} {\bibfnamefont {R.}~\bibnamefont
  {Coldea}}, \bibinfo {author} {\bibfnamefont {D.~A.}\ \bibnamefont {Tennant}},
  \bibinfo {author} {\bibfnamefont {E.~M.}\ \bibnamefont {Wheeler}}, \bibinfo
  {author} {\bibfnamefont {E.}~\bibnamefont {Wawrzynska}}, \bibinfo {author}
  {\bibfnamefont {D.}~\bibnamefont {Prabhakaran}}, \bibinfo {author}
  {\bibfnamefont {M.}~\bibnamefont {Telling}}, \bibinfo {author} {\bibfnamefont
  {K.}~\bibnamefont {Habicht}}, \bibinfo {author} {\bibfnamefont
  {P.}~\bibnamefont {Smeibidl}}, \ and\ \bibinfo {author} {\bibfnamefont
  {K.}~\bibnamefont {Kiefer}},\ }\href {\doibase 10.1126/science.1180085}
  {\bibfield  {journal} {\bibinfo  {journal} {Science}\ }\textbf {\bibinfo
  {volume} {327}},\ \bibinfo {pages} {177} (\bibinfo {year}
  {2010})}\BibitemShut {NoStop}%
\bibitem [{\citenamefont {Fava}\ \emph {et~al.}(2020)\citenamefont {Fava},
  \citenamefont {Coldea},\ and\ \citenamefont {Parameswaran}}]{fava2020glide}%
  \BibitemOpen
  \bibfield  {author} {\bibinfo {author} {\bibfnamefont {M.}~\bibnamefont
  {Fava}}, \bibinfo {author} {\bibfnamefont {R.}~\bibnamefont {Coldea}}, \ and\
  \bibinfo {author} {\bibfnamefont {S.~A.}\ \bibnamefont {Parameswaran}},\
  }\href@noop {} {\enquote {\bibinfo {title} {{Glide symmetry breaking and
  Ising criticality in the quasi-1D magnet CoNb$_2$O$_6$}},}\ } (\bibinfo
  {year} {2020}),\ \Eprint {http://arxiv.org/abs/2004.04169} {arXiv:2004.04169
  [cond-mat.str-el]} \BibitemShut {NoStop}%
\bibitem [{\citenamefont {Tippie}\ and\ \citenamefont
  {Clark}(1981{\natexlab{a}})}]{quinolinium1}%
  \BibitemOpen
  \bibfield  {author} {\bibinfo {author} {\bibfnamefont {L.~C.}\ \bibnamefont
  {Tippie}}\ and\ \bibinfo {author} {\bibfnamefont {W.~G.}\ \bibnamefont
  {Clark}},\ }\href {\doibase 10.1103/PhysRevB.23.5846} {\bibfield  {journal}
  {\bibinfo  {journal} {Phys. Rev. B}\ }\textbf {\bibinfo {volume} {23}},\
  \bibinfo {pages} {5846} (\bibinfo {year} {1981}{\natexlab{a}})}\BibitemShut
  {NoStop}%
\bibitem [{\citenamefont {Tippie}\ and\ \citenamefont
  {Clark}(1981{\natexlab{b}})}]{quinolinium2}%
  \BibitemOpen
  \bibfield  {author} {\bibinfo {author} {\bibfnamefont {L.~C.}\ \bibnamefont
  {Tippie}}\ and\ \bibinfo {author} {\bibfnamefont {W.~G.}\ \bibnamefont
  {Clark}},\ }\href {\doibase 10.1103/PhysRevB.23.5854} {\bibfield  {journal}
  {\bibinfo  {journal} {Phys. Rev. B}\ }\textbf {\bibinfo {volume} {23}},\
  \bibinfo {pages} {5854} (\bibinfo {year} {1981}{\natexlab{b}})}\BibitemShut
  {NoStop}%
\bibitem [{Note1()}]{Note1}%
  \BibitemOpen
  \bibinfo {note} {N. P. Armitage, private communication.}\BibitemShut {Stop}%
\end{thebibliography}%
\end{document}